\documentclass[a4paper,11pt]{article}

\usepackage{fullpage}

\usepackage{algorithm}
\usepackage{algpseudocode}

\usepackage{booktabs} % for formal tables
\usepackage{cite}

\usepackage{tikz}

\usepackage{latexsym}
\usepackage{url}
\usepackage{listings}
\usepackage{xspace}
\usepackage{color}

\usepackage{amsmath}
\usepackage{amssymb}
\usepackage{amsthm}

\newcommand{\eg}{e.g.\@\xspace}

\newcommand{\band}{\wedge}

\newcommand{\ceil}[1]{\lceil #1\rceil}

\newcommand{\mpisendrecv}{\textsf{MPI\_\-Sendrecv}\xspace}

\newcommand{\mpibarrier}{\textsf{MPI\_\-Barrier}\xspace}
\newcommand{\mpiscan}{\textsf{MPI\_\-Scan}\xspace}
\newcommand{\mpiexscan}{\textsf{MPI\_\-Exscan}\xspace}
\newcommand{\mpireducelocal}{\textsf{MPI\_\-Reduce\_local}\xspace}

\newcommand{\mpilong}{\texttt{MPI\_LONG}\xspace}
\newcommand{\mpibxor}{\texttt{MPI\_BXOR}\xspace}

\newcommand{\senddata}{\textsf{Send}\xspace}
\newcommand{\recvdata}{\textsf{Recv}\xspace}

\newcommand{\bidirec}[2]{\textsf{Send}(#1)\parallel\textsf{Recv}(#2)\xspace}

\newtheorem{theorem}{Theorem}

\newcommand{\gcc}{\texttt{gcc~12.1.0}\xspace}

\newcommand{\hydrampich}{mpich-4.1.2.\xspace}

\title{Communication Round and Computation Efficient Exclusive Prefix-Sums Algorithms (for \mpiexscan)}

\author{Jesper Larsson Tr\"aff\\
  TU Wien\\
  Faculty of Informatics\\
  Institute of Computer Engineering, Research Group Parallel Computing 191-4\\
  Treitlstrasse 3, 5th Floor, 1040 Vienna, Austria}
\date{July 2025}

\begin{document}

\maketitle

\begin{abstract}
  Parallel scan primitives compute element-wise inclusive or exclusive
  prefix sums of input vectors contributed by $p$ consecutively ranked
  processors under an associative, binary operator $\oplus$. In
  message-passing systems with bounded, one-ported communication
  capabilities, at least $\ceil{\log_2 p}$ or $\ceil{\log_2 (p-1)}$
  communication rounds are required to perform the scans.  While there
  are well-known, simple algorithms for the inclusive scan that solve
  the problem in $\ceil{\log_2 p}$ communication rounds with
  $\ceil{\log_2 p}$ applications of $\oplus$ (which could be
  expensive), the exclusive scan appears more
  difficult. Conventionally, the problem is solved with either
  $\ceil{\log_2 (p-1)}+1$ communication rounds (\eg, by shifting the
  input vectors), or in $\ceil{\log_2 p}$ communication rounds with
  $2\ceil{\log_2 p}-1$ applications of $\oplus$ (by a modified
  inclusive scan algorithm). We give a new, simple algorithm that
  computes the exclusive prefix sums in $q=\ceil{\log_2
    (p-1)+\log_2\frac{4}{3}}$ simultaneous send-receive communication
  rounds with $q-1$ applications of $\oplus$. We compare the three
  algorithms implemented in MPI against the MPI library native
  \mpiexscan primitive on a small, $36$-node cluster with a
  state-of-the-art MPI library, indicating possible and worthwhile
  improvements to standard implementations. The algorithms assume
  input vectors to be small so that performance is dominated by the
  number of communication rounds. For large input vectors, other
  (pipelined, fixed-degree tree) algorithms must be used.
\end{abstract}

\section{Introduction}

The parallel scan is a fundamental building block in parallel
algorithmics, often for bookkeeping and load balancing purposes, but
as often also directly in and for the algorithms themselves, see for
instance~\cite{Blelloch89,CopikGrosserHoeflerBientinesiBerkels22}.  Scan
operations compute the inclusive or exclusive prefix-sums over inputs
provided by $p$ successively ranked (numbered) processors under a
given, associative, binary operator $\oplus$. Each of the $p$
processors $r,0\leq r<p$ has an input vector $V_r$ of $m$ elements and
computes its element-wise, $r$th prefix sum $W_r$ as follows:
\begin{itemize}
\item
  Inclusive scan, for $0\leq r<p$:
  \begin{eqnarray*}
    W_r & = & \oplus_{i=0}^{r}V_i
  \end{eqnarray*}
\item
  Exclusive scan, for $0<r<p$:
  \begin{eqnarray*}
    W_r & = & \oplus_{i=0}^{r-1}V_i
  \end{eqnarray*}
\end{itemize}

For distributed memory, parallel message-passing programming, both
inclusive and exclusive scan operations are commonly standardized in
this way, \eg, in MPI~\cite{MPI-4.1} as \mpiscan and \mpiexscan,
respectively.

Arguably, the exclusive scan is the more important, more often used
variant of the scan primitives. In a shared-memory setting, where
input and output are stored in shared-memory arrays, the difference is
possibly not too important, since the $r$th exclusive prefix equals
exactly the $(r-1)$th inclusive prefix which can be read immediately in
the result array. In a distributed memory, message-passing setting,
this reduction from exclusive to inclusive scan involves communication
from processor $r-1$ to processor $r$ for all processors $r>0$.

Nevertheless, message-passing algorithms in the literature deal mostly
with the inclusive scan primitive, and trivially reduce the exclusive
scan primitive to an inclusive scan with one extra communication round
either after or before the inclusive scan (which need to be performed
on only $p-1$ processors).  As can easily be seen and as has been
discovered many times, the inclusive scan problem can be solved in
$\ceil{\log_2 p}$ communication rounds in which processors
simultaneously send and receive partial
results~\cite{HillisSteele86,KoggeStone73,KruskalRudolphSnir85}: We
will recapitulate the algorithm in Section~\ref{sec:algorithm}.  In systems
with one-ported communication capabilities, this is optimal, since the
last processor $r=p-1$ needs information (partial results) from all
$p$ processors and the number of processors from which $r$ can have
information after $k$ communication rounds is at most $2^k$. These
algorithms send and receive full, $m$-element input or partial result
vectors in each communication round, and are therefore mostly relevant
for vectors with small numbers of elements $m$.  For large vectors,
pipelined, fixed-degree tree algorithms should be used
\cite{Traff09:twotree,Traff06:scan}, which, however, require a larger
number of communication rounds and are therefore not as suitable for
small vectors. MPI libraries use different combinations of algorithms
and different choices for their implementations of \mpiscan and
\mpiexscan, but, as will be seen in Section~\ref{sec:evaluation},
often select logarithmic round, full-vector algorithms up to quite
large values of $m$.

In this short research note, we examine algorithms for the exclusive
scan operation, focusing on the number of communication rounds which
is one factor influencing the achievable, observable performance for
small vectors on real, distributed-memory parallel computing systems.
The reduction of the exclusive scan to an inclusive scan plus an extra
communication step will in the best case take $1+\ceil{\log_2 (p-1)}$
communication rounds, at least one more that the lower bound argument
gives. Modifying an inclusive scan algorithm to compute, in essence,
both inclusive and exclusive scans, leads to an algorithm taking
$\ceil{\log_2}$ communication rounds, but at the expense of two
applications of the $\oplus$ operator per round (except for the first
round, so the number of applications is actually $2\ceil{\log_2
  p}-1$). This can be a significant drawback when $\oplus$ is
expensive and as $m$ become larger. We present a new algorithm which
directly performs the exclusive scan in $q=\ceil{\log_2
  (p-1)+\log_2\frac{4}{3}}$ simultaneous send-receive communication
rounds with only $q-1$ applications of the $\oplus$ operator. This is
better than either of the other, sketched approaches. It remains open
whether an algorithm running in $\ceil{\log_2 (p-1)}$ rounds with a
most $\ceil{\log_2 (p-1)}$ applications of $\oplus$ exists.

We collect experimental evidence on a small high-performance cluster
with $36$ compute nodes, each with $32$ processor cores with a
commonly used, state-of-the-art MPI library to show that the
improvement, apart from being theoretically appealing, can be
worthwhile and have concrete, practical impact.

%\cite{HillisSteele86}
%\cite{KoggeStone73}
%\cite{KruskalRudolphSnir85}

\section{Algorithm}
\label{sec:algorithm}

In the logarithmic round inclusive scan algorithms, call them either
Hillis-Steele~\cite{HillisSteele86}, Kogge-Stone~\cite{KoggeStone73}
or Kruskal-Rudolph-Snir~\cite{KruskalRudolphSnir85}, the processors
maintain a partial result of the form
\begin{eqnarray*}
  W_r & = & \oplus_{i=\max(0,r-s_k+1)}^{r}V_i
\end{eqnarray*}
as an invariant that holds before round $k, k=0,1,\ldots,\ceil{\log_2
  p}-1$ for a sequence of straight doubling skips $s_k=2^k$.  Before
the first round with $k=0$, this invariant can easily be established
by locally copying $V_r$ into $W_r$, and if the invariant holds before
round $k$, processor $r$ can receive the computed partial result
$W_{r-s_k}$ from processor $r-s_k$ (as long as $r-s_k\geq 0$) and add
it with $\oplus$ to its own computed partial result $W_r$. Since
$s_k+s_k=2^k+2^k=2^{k+1}=s_{k+1}$, the invariant now holds before
round $k+1$. When $s_k>r$, the invariant implies that processor $r$
has indeed computed the $r$th inclusive prefix sum. Without intending
to unfairly give preference, we term this algorithm the
\emph{Hillis-Steele} or just \emph{straight doubling inclusive scan}
algorithm.

The doubling inclusive scan algorithm can be extended compute both the
inclusive and the exclusive scan by maintaining after the first
communication round instead the invariant
\begin{eqnarray*}
  W_r & = & \oplus_{i=\max(0,r-s_k+1)}^{r-1}V_i
\end{eqnarray*}
which implies that processor $r$ has computed the $r$th exclusive
prefix sum upon termination.  Processor $r$ would then have to send in
each round $W_r\oplus V_r$ to processor $r+s_k$ (as long as
$r+s_k<p$), requiring two applications of $\oplus$ in each round,
except the first.  We call this exclusive scan algorithm the
\emph{two-$\oplus$ doubling exclusive scan} algorithm.

The exclusive scan is apparently more elusive than the inclusive scan. We
would like to just maintain an invariant of the form
\begin{eqnarray*}
  W_r & = & \oplus_{i=\max(0,r-s_k)}^{r-1}V_i
\end{eqnarray*}
and use a similar, doubling communication pattern. Even with $s_0=1$, the
invariant does not hold before the first round, since processor
$r$ is missing the input from processor $r-1$ (for $r>0$). In order to
establish the invariant, we first let each processor $r$ receive
$V_{r-1}$ into $W_r$ from processor $r-1$.  For the next round,
the invariant now holds, and with $s_k=2^{k-1}$ for
$k=1,2,\ldots,\ceil{\log_2(p-1)}$, we can let each processor receive
the computed partial result $W_{r-s_k}$ from processor $r-s_k$ and add
it with $\oplus$ into the partial result $W_r$. Since
\begin{eqnarray*}
  W_{r-s_k} + W_r & = & 
  \left(\oplus_{i=\max(0,r-s_k-s_k)}^{r-s_k-1}V_i\right) \oplus
  \left(\oplus_{i=\max(0,r-s_k)}^{r-1}V_i\right) \\
  & = & \oplus_{i=\max(0,r-s_{k+1})}^{r-1}V_i
\end{eqnarray*}
the invariant is reestablished and holds again before round $k+1$. The
invariant is void for processor $r=0$, therefore after the initial
round with $s_0=1$, there is no further communication with processor
$r=0$ necessary, and $\ceil{\log_2 (p-1)}$ additional rounds suffice
to complete the exclusive scan. In total, the number of communication
rounds is $1+\ceil{\log_2 (p-1)}$. We term this exclusive scan
algorithm the \emph{$1$-doubling exclusive scan} algorithm. It is
essentially the same algorithm as first shifting the input from
processor $r$ to processor $r+1$ and then performing a straight
doubling scan on the $p-1$ processors $r,r>0$. This algorithm has the
advantage that it performs only $\ceil{\log_2 (p-1)}$ applications of
$\oplus$.

To finally get the number of communication rounds closer to only
$\ceil{\log_2 (p-1)}$ without any additional applications of $\oplus$,
that is, eliminating the constant term, a new idea is needed. After
the initial round which establishes the invariant with $W_r=V_{r-1}$,
we observe that if we let processor $r$ receive from processor $r-2$
with $s_1=2$ the partial result $W_{r-2}\oplus V_{r-2}$,
processor $r$ can now compute
\begin{displaymath}
  \left(W_{r-2}\oplus V_{r-2}\right) \oplus W_r  =
  V_{r-3}\oplus V_{r-2}\oplus V_{r-1} 
\end{displaymath}
which is exactly the desired invariant situation if we take
$s_2=3$. After this, for all following rounds, $k=2, 3,\ldots$, we
take $s_k=3\cdot 2^{k-2}$. The doubling rounds continue as long as
$(p-1)-3\cdot 2^{k-2}>0$, that is $k-2\geq \log_2\frac{p-1}{3}$. This
gives to $\ceil{\log_2(p-1)+\log_2\frac{4}{3}}$ required communication
rounds.  We term this exclusive scan algorithm the
\emph{$123$-doubling exclusive scan} algorithm.

\begin{algorithm}
  \caption{The $123$-doubling exclusive scan algorithm for processor
    $r, 0\leq r<p$ with skips $s_0=1,s_1=2$ and $s_k=3\cdot 2^{k-2}$
    for $k>1$.  Each processor has input in $V$ and computes the $r$th
    element wise exclusive prefix sum in rank order into $W$. The
    associative reduction operator is $\oplus$.}
  \label{alg:exscan}
  \begin{algorithmic}
    \Procedure{ExScan}{$V,W,\oplus$}
    \State $t,f\gets r+s_0,r-s_0$ \Comment To- and from-processors for round $0$
    \If{$0\leq f\band t<p$}
    \State $\bidirec{V,t}{W,f}$
    \ElsIf{$t<p$}
    \State $\senddata(V,t)$
    \ElsIf{$0\leq f$}
    \State $\recvdata(W,f)$
    \EndIf
    \State $t,f\gets r+s_1,r-s_1$ \Comment To- and from-processors for round $1$
    \If{$0\leq f\band t<p$}
    \State $W'\gets W\oplus V$
    \State $\bidirec{W',t}{T,f}$
    \State $W\gets T\oplus W$
    \ElsIf{$t<p\band r=0$}
    \State $\senddata(V,t)$
    \State \Return \Comment Processor $r=0$ done
    \ElsIf{$t<p\band r>0$}
    \State $W'\gets W\oplus V$
    \State $\senddata(W',t)$
    \ElsIf{$0\leq f$}
    \State $\recvdata(T,f)$
    \State $W\gets T\oplus W$
    \EndIf
    \State $k\gets 2$
    \State $t,f\gets r+s_k,r-s_k$
    \Comment To- and from-processors for remaining rounds
    \While{$0<f\band t<p$}
    \State $\bidirec{W,t}{T,f}$
    \State $W,k\gets T\oplus W,k+1$
    \State $t,f\gets r+s_k,r-s_k$ 
    \EndWhile
    \While{$t<p$}
    \State $\senddata(W,t)$
    \State $k\gets k+1$
    \State $t\gets r+s_k$ \Comment To-processor
    \EndWhile
    \While{$0<f$}
    \State $\recvdata(T,f)$
    \State $W,k\gets T\oplus W,k+1$
    \State $f\gets r-s_k$ \Comment From-processor
    \EndWhile
    \EndProcedure
  \end{algorithmic}
\end{algorithm}

The $123$-doubling exclusive scan algorithm is shown in full
detail as Algorithm~\ref{alg:exscan}. The notation $\bidirec{W,t}{T,f}$
denotes a simultaneous send and receive operation with data in buffers
$W$ and $T$, to and from processors $t$ and $f$. Summarizing the
explanation from above, we have the following theorem.

\begin{theorem}
  \label{thm:circulantscan}
  The exclusive scan problem is solved $q=\ceil{\log_2
    (p-1)+\log_2\frac{4}{3}}$ simultaneous send-receive communication
  rounds with $q-1$ applications of the associative, binary operator
  $\oplus$ by Algorithm~\ref{alg:exscan}.
\end{theorem}

\section{Experiments}
\label{sec:evaluation}

\begin{figure}
  \begin{center}
    \includegraphics[width=.45\textwidth]{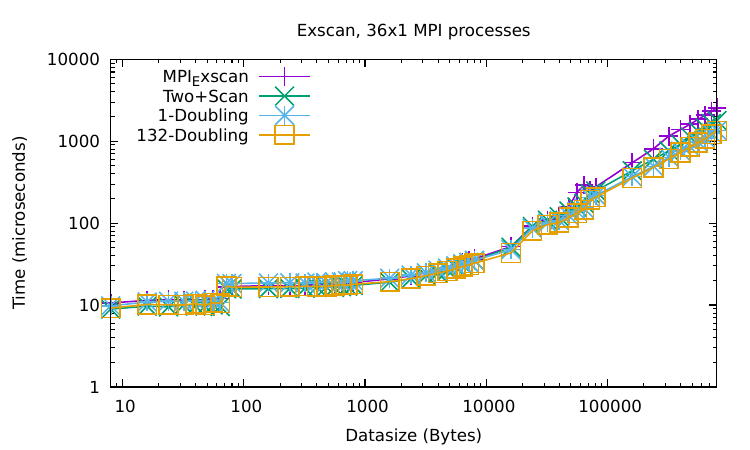}
    \includegraphics[width=.45\textwidth]{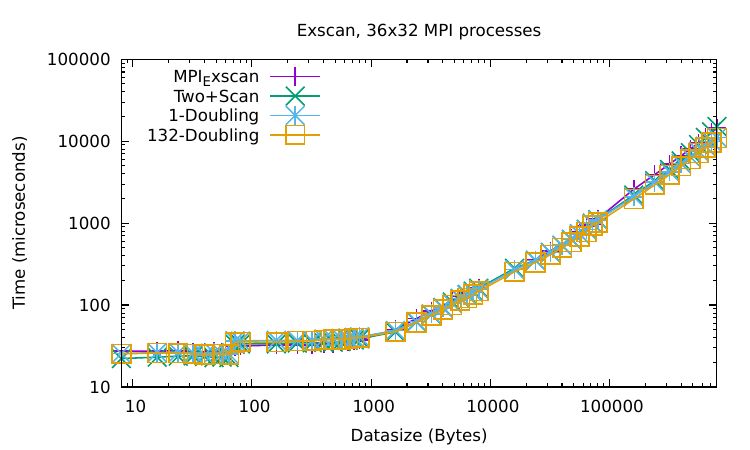}
  \end{center}
  \caption{Results for the \mpiexscan operation, natively with
    \mpiexscan, the with two-$\oplus$ doubling, the $1$-doubling, and
    the new $123$-doubling algorithm shown as Algorithm~\ref{alg:exscan}
    with \hydrampich in $p=36\times 1$ and $p=36\times 32$
    % $p=25\times 1$ and $p=25\times 32$
    MPI process configurations. Times are in microseconds as a
    function of number of bytes in input and output.  Both axes are
    logarithmic.}
  \label{fig:exscan36}
\end{figure}

We have implemented the three exclusive scan algorithms described in
Section~\ref{sec:algorithm}, in particular Algorithm~\ref{alg:exscan},
in MPI~\cite{MPI-4.1} using standard operations. Simultaneous send and
receive is implemented with \mpisendrecv, and local application of a
pre- or user-defined MPI operator with \mpireducelocal. The latter is
a two-argument operation, taking an input and an input-output vector
and reducing these together in this order. For
Algorithm~\ref{alg:exscan} in communication round one, where an
intermediate, partial result $W_{r-1}\oplus V_r$ has to be sent, a
three-argument local reduction function would have been convenient and
desirable~\cite{Traff23:attributes}. We aim to estimate where the
three algorithms differ and how they compare to the algorithm
implemented in the chosen MPI library for \mpiexscan.

Our experimental system is a medium sized $36\times 32$ processor
cluster with $36$~dual socket compute nodes, each with two Intel(R)
Xeon(R) Gold 6130F $16$-core CPUs. The compute nodes are
interconnected via dual Intel Omnipath interconnects each with a
bandwidth of $100$ Gbytes/s.  We report results only with the
apparently best performing MPI library available to us (with respect
to the \mpiscan and \mpiexscan collective operations) which is the
\hydrampich library, and observe that MPI libraries differ both in
concrete performance and in scalability characteristics.  The
implementations and benchmarks were compiled with \gcc with the \texttt{-O3}
option.

As element type for the scan operations, we have used \mpilong, and
\mpibxor as the binary operator. Our benchmarking procedure performs,
for each element count, 200 repetitions with 15 warmup
measurements. We synchronize the MPI processes with \mpibarrier
(twice), and for each experiment determine the time for the slowest
process to complete the exclusive scan operation. Over the 200
repetitions, the minimum of these times is reported and
plotted~\cite{Traff12:mpibenchmark}. The results for element counts
from $0$ to $100\,000$ in $p=36\times 1$ and $p=36\times 32$ MPI
process configurations are plotted in Figure~\ref{fig:exscan36}.

\begin{table}
  \caption{Measured running times for the native \mpiexscan operation
    and the three algorithms with the \hydrampich library in the
    $p=36\times $ and $p=36\times 32$ MPI process configurations,
    respectively.}
  \label{tab:sometimes}
  \begin{center}
    \begin{tabular}{rrrrr}
      & \multicolumn{4}{c}{$p=36\times 1$ MPI processes} \\
      \midrule
      $m$ \mpilong &
      \mpiexscan & two-$\oplus$ doubling & $1$-doubling & $123$-doubling \\
      & ($\mu$seconds) & ($\mu$seconds) & ($\mu$seconds) & ($\mu$seconds) \\
      \toprule
      1
& 10.61 
& 8.92 
& 9.79 
& 9.17 
      \\
      10
& 16.86
& 15.68
& 18.29
& 16.58
      \\
      100
& 18.78
& 17.34
& 19.83
& 17.95
      \\
      1000
& 36.77
& 34.98
& 35.13
& 32.38
      \\
      10\,000
& 276.31
& 247.39
& 218.06 
& 207.29 
      \\
      100\,000
& 2558.52 
& 1789.40 
& 1351.72 
& 1333.91 
      \\
      \bottomrule
      & \multicolumn{4}{c}{$p=36\times 32$ MPI processes} \\
      \midrule
      $m$ \mpilong &
      \mpiexscan & two-$\oplus$ doubling & $1$-doubling & $123$-doubling \\
      & ($\mu$seconds) & ($\mu$seconds) & ($\mu$seconds) & ($\mu$seconds) \\
      \toprule
      1
& 27.27
& 22.23
& 25.61
& 25.36
      \\
      10
& 31.59
& 33.55
& 36.36
& 35.67
      \\
      100
& 37.55
& 38.77
& 40.96
& 39.97
      \\
      1000
& 160.34
& 160.40
& 155.99
& 147.20
      \\
      10\,000
& 1124.82
& 1103.67
& 1095.03
& 1018.43
      \\
      100\,000
& 14456.12
& 15107.82
& 11120.00
& 10921.26
      \\
      \bottomrule
    \end{tabular}

  \end{center}
\end{table}
    
For configurations with one MPI process per compute node
(\eg, $p=36\times 1$), the difference between the
algorithms is particularly clear and significant. For, say,
$m=10\,000$ elements, the library native \mpiexscan implementation
takes $276 \mu s$, and the best, $123$-doubling algorithm only $207
\mu s$. This is an improvement by $25\%$. The two other algorithms are
in between. In particular, the $1$-doubling exclusive scan is
sometimes on par with the $123$-doubling scan algorithm, but never
better, and often noticeably slower. This can be seen in the
$p=36\times 32$ MPI process configuration, where the times for
$m=1000$ elements for the four algorithms are $1124, 1103, 1095$ and
$1014$ $\mu s$, respectively. The actual running times are shown in
Table~\ref{tab:sometimes} which more clearly shows the difference
between the three algorithms. For small number of processes, the
additional communication round incurred by the $1$-doubling algorithm
affects performance. As the number of elements $m$ grows, and for
larger MPI process counts, the additional application of the $\oplus$
operator per round by the two-$\oplus$ doubling algorithm has a
negative effect. For very small $m$, this algorithm is sometimes the
best, however.

In summary, the \mpiexscan implementation in the \hydrampich library
can be significantly improved, with the most improvement by the new,
$123$-doubling algorithm.

\section{Summary}

In this research note, we surveyed three direct algorithms for the
message-passing exclusive scan operation, and in particular gave a new
algorithm that computes the exclusive scan in $q=\ceil{\log_2
  (p-1)+\log_2\frac{4}{3}}$ simultaneous send-receive communication
rounds and $q-1$ applications of the given, associative (and possibly
expensive) binary operator. An experimental study indicates that the
improvements are worthwhile for the implementation of \mpiexscan for
MPI libraries. It is an open question whether an algorithm exists that
can compute the exclusive scan in $\ceil{\log_2 (p-1)}$ communication
rounds and at most the same number of application of the operator.

\bibliographystyle{plainurl}
\bibliography{traff,parallel} 

\end{document}